\documentclass[iop]{emulateapj}
\usepackage{verbatim}
\usepackage{mathtools}
\usepackage{natbib}
\usepackage{bm}        
\usepackage{amssymb}   
\usepackage{booktabs}

\begin{document}
\title{Deriving galaxy cluster velocity anisotropy profiles \\ from a joint analysis of dynamical and weak lensing data}

\author{ Alejo Stark\altaffilmark{1},Christopher J. Miller\altaffilmark{1,2}  Vitali Halenka \altaffilmark{2}}
\altaffiltext{1}{Department of Astronomy, University of Michigan, Ann Arbor, MI 48109 USA}
\altaffiltext{2}{Department of Physics, University of Michigan, Ann Arbor, MI 48109, USA}
\email{alejo@umich.edu}


\begin{abstract}
We present an analytic approach to lift the mass-anisotropy degeneracy in clusters of galaxies by utilizing the line-of-sight velocity dispersion of clustered galaxies jointly with weak lensing-inferred masses. More specifically, we solve the spherical Jeans equation by assuming a simple relation between the line-of-sight velocity dispersion and the radial velocity dispersion and recast the Jeans equation as a Bernoulli differential equation which has a well-known analytic solution. We first test our method in cosmological N-body simulations and then derive the anisotropy profiles for 35 archival data galaxy clusters with an average redshift of $\langle z_c \rangle = 0.25 $. The resulting profiles yield a weighted average global value of $\langle \beta( 0.2 \leq R/R_{200} \leq 1 )\rangle = 0.35 \pm 0.28$ (stat) $\pm 0.15$ (sys). This indicates that clustered galaxies tend to globally fall on radially anisotropic orbits. We note that this is the first attempt to derive velocity anisotropy profiles for a cluster sample of this size utilizing joint dynamical and weak lensing data. 
\end{abstract}



\section{Introduction}
\label{sec:intro}
Clusters of galaxies are fruitful sites of experimentation that allow us to unravel the nature of gravitation, understand the ingredients of our cosmological models (such as dark matter and dark energy), as well as make sense of the complex astrophysical processes involved in both the evolution of galaxies and the intracluster medium \citep{KravtsovBorgani}. From the standard model of structure formation, galaxy clusters are  both the last and the largest structures to have formed in the history of cold dark matter halos. These halos emerged from small seeds of density perturbations embedded in the expanding Hubble flow that have grown hierarchically since then \citep{GunnGott1972,Bertschinger1985,filmore1984}. 

Results from cosmological N-body simulations of cold dark matter halos have shown that the mass profile of clusters can be universally characterized by simple models, such as the NFW (Navarro-Frenk-White) or the Einasto profiles \citep{einasto65,navarro97}. Observationally, this has meant that mass estimation techniques, such as those depending on weak lensing shear measurements, utilize these universal profiles to infer cluster masses. While most of the observed cluster weak lensing mass profiles are derived using the NFW profiles (as an example, see the metacatalog compiled in \citep{sereno}), recent work using high resolution simulations has shown that an Einasto model, combined with a transition to a steep power-law beyond $\sim R_{200}$ (with respect to the critical density of the Universe), provides the most accurate description of density profile for clusters over a wide range of accretion histories and redshifts \citep{Diemer14,Diemer2015,More15,More16}. 

At the same time, the internal dynamics of galaxy clusters continues to be a fruitful avenue of research. Since the seminal investigations of the ``nebulae'' within the Coma cluster by \cite{zwicky}, clusters have long been observational laboratories used to characterize the dynamical evolution of gravitationally-bound objects. In early N-body simulations, the velocity dispersion of the cluster particles was identified as a simple yet powerful way to infer the gravitational potential \citep{Evrard1986}. As the resolution of the N-body simulations increased to include sub-halos as representative of the galaxy populations, the 3-dimensional velocity dispersion was better characterized, including the effects of sub-halo sampling bias \citep{Evrard2008,Wu2013,gifford2013,giffordBias, Saro2013}. Through the proliferation of large-scale spectroscopic surveys and  multi-object spectroscopic instrumentation, we are evidencing the proliferation of precise dynamical data of galaxy clusters \citep{Miller2005, Becker2007,Farahi16, Bayliss2017}

However, despite this proliferation of data, a dynamical quantity that is yet not well characterized is the galaxy cluster velocity anisotropy profile $\beta$  \citep{BinneyTremaine1987},
\begin{equation}
\beta = 1 - \frac{\sigma_t^2}{\sigma_r^2},
\label{eq:betasimple}
\end{equation}
where $\sigma_t^2$ and $\sigma_r^2$ signify the tangential and radial velocity dispersions, respectively, of galaxies at a given radial bin. As an example, in the case of a totally isotropic velocity distribution of the galaxies, Eq. \ref{eq:betasimple}  yields $\beta = 0$, whereas a totally radial velocity distribution yields $\beta = 1$. While the upper limit of the parameter is unity, the lower limit can in principle be $\beta = -\infty $. Values of $\beta$ below 0 entail tangential galactic orbits.

Typically, inferring velocity anisotropy profiles follows the long tradition of studies that have modeled galaxy clusters as collisionless systems described by the anisotropic Jeans equation \citep{BinneyTremaine1987,BinneyMamon1982,Solanes1990},
\begin{equation} 
\frac{d(\rho \sigma_r^2)}{dr} + \frac{2\beta \sigma_r^2 \rho}{r} =  -\rho \frac{d\phi}{dr},
\label{eq:jeans}
\end{equation} 
where the potential-density pair of the system is defined respectively by the quantities $\phi$ and $\rho$, and $\sigma_{r}^2$ represents the radial velocity dispersion of the tracers (in our case, galaxies). 

In the Jeans equation, $\beta$ is degenerate with the mass profile through both the density and the derivative of the potential. This is the so called ``mass-anisotropy degeneracy" \citep{Merritt1987}. One obvious solution to this degeneracy is to attain an independently measured mass \citep{BinneyMamon1982,Solanes1990}. Previous work in this area has provided estimates for cluster anistropy profiles by using X-ray masses \citep{host2008,Hwang,benatov}.  Other authors have attempted to break the mass-anisotropy degeneracy using just the dynamical information in the cluster phase-spaces \citep{wojtak,lokasabell,BivianoKatgert2004}. However, there are very few examples of higher precision measurements of the anisotropy profile using large samples of clusters \citep{host2008,wojtak}. 

Given the proliferation of data and techniques to characterize the mass profiles of galaxy clusters,  various papers have focused deriving $\beta$ for individual clusters by combining mass profiles inferred through different techniques  \citep{macs1206, Aguerri2017,Annunziatella2016,Munari2014,lemze2009}.  Most of these studies conclude that while the orbits of clustered galaxies are unlikely to be tangential, the overall scatter is often too large to determine either the degree of isotropy or radial anisotropy of galactic orbits that is expected from the results of N-body simulations -- which tends to show only a weak radial anisotropy within the virial radius \citep{iannuzzi,lemze2009,serra}. In contrast to these individual cluster studies, we are more interested in deriving a ``global" galaxy anisotropy profile from a relatively large sample size of clusters, in the spirit of work such as that of \cite{wojtak}, which derived velocity anisotropy profiles for 41 clusters.

In what follows, we derive an average anisotropy profile for 35 galaxy clusters. We follow the intuition of (but also take a different approach than)  \cite{Natarajan97}--a seminal paper that is the first to attempt to do a joint dynamics-weak lensing constraint of $\beta$ using the Jeans equation. We carefully test our approach using synthetic clusters produced through the N-body Millennium simulations \citep{springel2005millenium}. We then apply our new algorithm to archival data from 35 galaxy clusters with both weak lensing mass profiles and extensive spectroscopic coverage. In  contrast to  \cite{Natarajan97}, which largely focuses on characterizing the degree of anisotropy of the core of a single cluster, we derive profiles for our 35 clusters that extend out to one virial radii and also sidestep the complications of modeling galaxy cluster cores. In particular, we find that within 0.2 - 1 virial radii, the average galaxy cluster velocity anisotropy profile of the 35 clusters derived from archival data tends to be radially anisotropic with a small statistical scatter. In what follows, we attempt to address how these results compare to other derivations of $\beta$ and discuss its implications.  We highlight that this is the first attempt to derive a global $\beta$ profile of galaxy clusters with joint weak lensing and dynamical data. Lastly, we note that when using the Jeans equation (Eq. \ref{eq:jeans}) we are assuming that galaxies are distributed like the underlying diffuse matter, represented by density profile $\rho$. As shown by \cite{Munari2014} this may not be the case for all systems. As such, one may opt to use the number density of the tracers (i.e. the galaxy surface density). However, we opt against this approach because to use this observable successfully we would have to correct for spectroscopic incompleteness-something we cannot currently accomplish with the archival data set we utilize. 

The outline of our paper is as follows: in Section \ref{sec:data} we describe the data we used to both test our approach in cosmological N-body simulations, as well as the joint weak lensing and spectroscopic archival data of the 35 galaxy clusters; in Section \ref{sec:TheoryObs} we clarify the observables derived from the aforementioned data and then describe how they are used to derive $\beta$ profiles in Section \ref{sec:theory_exp}; Section \ref{sec:results} presents the results of our approach; lastly, in Section \ref{sec:discussion} we discuss our results in light of other derivations of $\beta$ as well as the systematics affecting our probe. For the case of synthetic data (real data)   we assume a flat $\Lambda$CDM cosmology with $\Omega_{M}= 0.25 (0.3)$ , $\Omega_{\Lambda}=1-\Omega_{M}$, and $H_0 = 100 h \text{ km}\text{ s}^{-1}\text{ Mpc}^{-1} $ with $h = 0.73 (0.7).$


\section{data}
\label{sec:data}

In this section we describe the synthetic data we used to test our approach with N-body simulations, and then describe the archival of the 35 galaxy clusters that contain both weak lensing and spectroscopic observations.

\subsection{N-body simulations}

We test our method on dark matter halos generated by the N-body cosmological simulations of the Millennium simulation (MS)  \citep{springel2005millenium}. In particular, we use the particle data from these simulations to measure the cluster Einasto density profiles. We treat these density profiles as the ``weak lensing'' data, since they trace the underlying matter distribution.

More specifically, we select 100 halos from the Millennium Simulation to test our approach. Our sample aims for fair cluster mass sampling over the range $\sim 10^{14} - 10^{15} M_{\odot}$. The average mass of our sample is $\langle M \rangle = 2.34 \times 10^{14} M_{\odot}$ and the average critical radius is $\langle R_{200} \rangle = 0.95$ Mpc. We then use the semi-analytic galaxy catalogs from \cite{guo10} to define our projected cluster radius/velocity phase-spaces. These phase-spaces typically contain between 100-200 galaxies within $R_{200}$ and $\pm 3000$ km s$^{-1}$, which is similar to the observed data (see next sub-section). The volume around each halo is a cube with box length 60 Mpc h$^{-1}$. This box length is large enough to make line-of-sight projections to include realistic phase-space contamination from typical galaxy peculiar velocities (i.e., $\pm \sim 3000$ km s$^{-1}$). We calculate the line-of-sight velocity dispersions using 100 random orientations around each cluster. We then use these to measure the median line-of-sight velocities $v_{los}$ as well as an error based on the 1$\sigma$ scatter of the individual profiles. We explain this more thoroughly in the next sections.

\subsection{Archival data}

To derive the global $\beta$ profile in real data we use data from 35 clusters found in the literature. To find the 35 archival data clusters we used the VizieR catalog \citep{vizier} to search for redshifts of the galaxy clusters (\cite{hecs, Maurogordato2008, Owers2011122O, Owers201127O, Tyler2013, Girardi2008, Oemler2009, Agulli2016, Tran2007, Demarco2010, Lemze2013, Moran2007, Geller2014, Girardi2015, Edwards2011}) that also have weak lensing data (\cite{hoekstra2015, okabe2008, Okabe2010, okabe2015, Umetsu2015, Cypriano2004, Pedersen2007, Medezinski2016, Foex2012, Clowe2000, Jee2011, Smail1997}). We tabulate this joint dynamic-weak lensing data set on Table \ref{table1}. Note that while we cite the original papers,  the weak lensing masses (and their respective errors) we use in our analysis were taken from the standardized \cite{sereno} meta catalog.

More specifically, we note that 10 clusters in our sample have $57 < N < 100$ galaxies while the remaining 25 clusters have more than 100 galaxies within $ R_{200}$ and within their escape velocity profile ($v_{esc}$). The mass range of the archival data lies between $4.1\times10^{14}M_{\odot}$ and $2.06\times10^{15}M_{\odot}$. Note that the meta catalog only lists masses inferred from NFW fits to weak lensing shear measurements.  As is detailed in the next section, we transform the NFW fit parameters to those of the Einasto model.

\begin{table}[]
\centering
\caption{List of Galaxy Clusters and References}
\label{table1}
\begin{tabular}{@{}llll@{}}
\toprule
Cluster name\footnote{While we cite the original papers above, the weak lensing masses (and their respective errors) we use in our analysis were taken from the \cite{sereno} meta catalog. More specifically, \cite{sereno} standardizes the $M_{200}$  masses for the clusters shown above (as inferred from each reference listed in the ``weak lensing" column) for the fiducial cosmology mentioned in our introduction.} & Redshift & Weak lensing\footnote{The abbreviations in this column refer to the following papers: H15= \cite{hoekstra2015}, O08 = \cite{okabe2008}, O10 = \cite{Okabe2010}, O15= \cite{okabe2015}, A14 = \cite{Applegate2014}, U15= \cite{Umetsu2015}, C04 = \cite{Cypriano2004}, P07 = \cite{Pedersen2007}, M16 = \cite{Medezinski2016}, F12 = \cite{Foex2012}, CL00 = \cite{Clowe2000}, J11 = \cite{Jee2011}, S97 = \cite{Smail1997}. We averaged over multiple weak lensing sources to get $M_{200}$ as well as the errors of the clusters A2163 and A2219.} & Galaxy redshifts \\ \midrule
A1682	&	0.227	&	P07	&	Rines et al. '13	\\
A1553	&	0.167	&	C04	&	Rines et al. '13	\\
A1423	&	0.214	&	O15	&	Rines et al. '13	\\
A2163	&	0.201	&	H15/R08	&	Maurogordato et al. '08	\\
A2034	&	0.113	&	O08	&	Rines et al. '13	\\
A2029	&	0.077	&	C04	&	Tyler et al. '13	\\
A2009	&	0.152	&	O15	&	Rines et al. '13	\\
A2219	&	0.226	&	O10/O15/A14	&	Rines et al. '13	\\
A2744	&	0.306	&	M16	&	Owers et al. '11b	\\
A520	&	0.201	&	H15	&	Girardi et al. '08	\\
A851	&	0.405	&	F12	&	Oemler et al. '09	\\
A85	&	0.055	&	C04	&	Agulli et al. '16	\\
A773	&	0.217	&	O15	&	Rines et al. '13	\\
ZwCl3146	&	0.289	&	O15	&	Rines et al. '13	\\
BLOXJ1056	&	0.831	&	CL00	&	Tran et al. '07	\\
RXJ1720	&	0.160	&	O15	&	Owers et al. '11a	\\
RXJ0152	&	0.837	&	J11	&	Demarco et al. '10	\\
RXCJ1504	&	0.217	&	O15	&	Rines et al. '13	\\
A2111	&	0.229	&	H15	&	Rines et al. '13	\\
A611	&	0.287	&	O10	&	Lemze et al. '13	\\
ZwCl0024	&	0.395	&	S97	&	Moran et al. '07	\\
A2259	&	0.161	&	H15	&	Rines et al. '13	\\
A1246	&	0.192	&	H15	&	Rines et al. '13	\\
A697	&	0.281	&	O10	&	Rines et al. '13	\\
A1689	&	0.184	&	U15	&	Rines et al. '13	\\
A1914	&	0.166	&	H15	&	Rines et al. '13	\\
A1835	&	0.251	&	H15	&	Rines et al. '13	\\
A267	&	0.229	&	O15	&	Rines et al. '13	\\
A1763	&	0.231	&	H15	&	Rines et al. '13	\\
A963	&	0.204	&	O15	&	Rines et al. '13	\\
A383	&	0.189	&	O15	&	Geller et al. '14	\\
A2142	&	0.090	&	O08	&	Owers et al. '11a	\\
RXCJ2129	&	0.234	&	O15	&	Rines et al. '13	\\
MACS1206	&	0.440	&	F12	&	Girardi et al. '15	\\
Coma & 0.0231 & H15 & Edwards, Fadda'11 \\

\bottomrule
\end{tabular}
\end{table}


\section{Theoretical Observables}
\label{sec:TheoryObs}
Before we can derive $\beta$ profiles for our set of both synthetic data and the 35 clusters, we extract both the dynamical data of clustered galaxies thorough their line-of-sight velocities as well as their respective cluster mass profiles. In the following two sub-sections we describe the relevant observables and other quantities that we use in our approach to derive $\beta$.

\subsection{Galaxy line-of-sight velocity dispersion profiles}
The redshifts of galaxies along a line-of-sight are used to generate the line-of-sight velocities vs. projected distance ($v_{los}$ vs. $R$) ``phase spaces." From these phase spaces we infer the line-of-sight velocity dispersion. In particular, for a given cluster, given some angular separation from the cluster center ($\theta$) the line-of-sight ($v_{los}$) velocities at the cluster's redshift ($z_c$)  can be inferred via,
\begin{equation} 
v_{los} =  c\frac{(z - z_c)}{{(1+z_c)} },
\end{equation} 
where $c$ denominates the speed of light, and  $z$ is the redshift of the individual galaxies along the line-of-sight. Now, the projected distance from the cluster's center ($R$) can be inferred from the angular diameter distance ($d_A$) and the aforementioned angular separation ($\theta$), 
\begin{equation} 
R = d_{A}(z) \theta = \bigg[\frac{1}{1+z} \frac{c}{H_0} \int_0^z \frac{dz'}{E(z')}\bigg] \theta,
\end{equation} 
where the cosmological evolution of the energy densities for the aforementioned $\Lambda$CDM cosmology is given by  $E(z) =  \sqrt{ (1- \Omega_{M}) + \Omega_{M} (1+z)^3 }$. Note that for the cluster center and cluster redshift $z_c$ we use what has been specified for each respective cluster in the spectroscopic catalog described in the previous section.

Having created a phase space for a given cluster, we then calculate the velocity dispersion profile of the cluster. However, before we can do this we remove galaxies that may exist in our phase space simply because they are observed within our line-of-sight, but do not exist within the virial sphere of the cluster. If these ``interloper" galaxies are not removed, the resulting velocity dispersion profiles will be biased. To do this, we apply the shifting-gapper technique of \cite{gifford2013} (see further references about interloper removal techniques therein, as well as a detailed statistical and systematic analysis of this technique).

Having identified and removed interlopers from the cluster phase space, we bin the galaxies radially to calculate the line-of-sight velocity dispersion profile. More specifically, for each radial bin, we calculate the line-of-sight dispersion via the following robust estimator,
\begin{equation}
\sigma_{los}^2= \sum_j \frac{\langle v_{los}(r_j)^2 \rangle}{N_{gal,j}-1},
\label{eq:sigmavlos}
\end{equation}
where we sum over the number of galaxies $N_{gal,j}$ within a given radial bin  $r_j$ with width $\Delta R$.  $j$ refers to the individual galaxies within that bin. We attain a radial profile by  calculating the dispersion of the various radial bins of a given cluster.  We calculate the error on each radial bin via a bootstrapping re-sampling algorithm, and find that the error on any given radial bin is on average $\sim 50$ km s$^{-1}$.

Having inferred the line-of-sight velocity profile we then fit it with a simple power law, \citep{Carlberg97,Aguerri2017}

\begin{equation}
\sigma_{f}(r)= \sigma_{f,0} (1+R)^{p} ,
\label{eq:sigmafit}
\end{equation}
where the central projected velocity dispersion is given by $\sigma_{f,0}$ and $p$ signifies the exponent.

For the case of the synthetic data we use a radial bin width of $\Delta R = 0.2$ Mpc and we fit Eq. \ref{eq:sigmafit} to the $\sigma_{los}$ profiles for the radial range $ 0.1 \leq R/R_{200} \leq 1.5 $ Mpc. We find that the fits match the measured $\sigma_{los}$ profiles with no bias and to high precision.

Clusters in the archival data are occasionally less well-sampled than in the simulation data. Therefore,  we define the center of each radial bin after first ensuring that there is 20 galaxies within it. We take the mean radius of these twenty galaxies in each bin. As such, for a given cluster in the synthetic data, the radial bin width can range from $\Delta r \sim 0.1$ to 0.2 Mpc. Note that in contrast to the synthetic clusters, we fit Eq. \ref{eq:sigmafit} over the radial range for all the data available, which varies in spatial extent depending on the cluster. We find that, on average, for the entire synthetic sample, $\sqrt{\langle (\sigma_{los}^{fit} - \sigma_{los})^2 \rangle} \sim 153$ km s $^{-1}$. Furthermore, also for the entire synthetic sample, we find that the averaged difference between the fits and the measured line-of-sight velocity profiles is $\langle \sigma_{los}^{fit} - \sigma_{los} \rangle \sim -43$ km s $^{-1}$. This bias is at about the level of the individual errors on the galaxy velocities. Note that this stipulates that, on average, our fits are slightly biased low compared to the noisy data. This effect is due to the bias-variance trade-off and indicates that there is likely a better fitting function that could be applied. We also note that, as reported in \cite{giffordBias}, improving the completeness of the cluster phase spaces would improve these fits. Nonetheless, we note that they are within the statistical error. We further discuss the implications of our $\sigma_{los}$ fits on Section \ref{sec:discussion}.

\subsection{Galaxy cluster mass profiles}
The other observable of interest  is the mass profile of galaxy clusters. For the case of the N-body simulations we attain the parameters describing these profiles by fitting the dark matter halo with an Einasto profile. The Einasto representation of the dark matter halo density profile \citep{einasto65} is a three parameter model ($n, \rho_{0}, r_0$ -- which represent the index, the normalization density, and the scale radius, respectively) described by the following fitting formula for the density profile,

\begin{equation}
\rho(r) =  \rho_0 \exp \Bigg[ -\bigg(\frac{r}{r_0}\bigg)^{1/n} \Bigg].
\label{eq:einasto_den}
\end{equation}
For the case of the archival data, however, we utilized the weak lensing parameters listed in the catalog for each respective cluster as described in the previous section (see Table \ref{table1}). Because the vast majority of  characterizations of galaxy cluster's density profiles are represented in terms of NFW fits, we fit the Einasto profile to the NFW density profile for each respective cluster in the radial range $ 0.05 \leq R \leq R_{200}$ Mpc and ensure that they fit to high precision within that radial range as expected by \cite{sereno2016}. To do this, however, we must also use a mass-concentration relation from the cluster metacatalog, where the concentration ($c_{200}$) as a function of mass ($M_{200}$) and cluster redshift ($z_c$) is given by \citep{sereno},
 \begin{equation}
 c_{200}(M_{200}, z_c) = A_{200} \Bigg(\frac{M_{200}}{M_{piv}} \Bigg)^{B_{200}} (1+z_c)^{C_{200}}.
 \label{eq:Mcrelation}
 \end{equation}
  Where $A_{200}= 5.71$, $B_{200} = -0.084$,  $C_{200} = -0.47$ and $M_{piv} = 2\times 10^{12} h^{-1} M_{\odot}$. As usual, the quantities with subscript ``200" are defined in terms of  the radius $R_{200}$ which is the distance at which the density enclosed within a sphere drops to 200 times the critical density of the Universe. The mass $M_{200}$ is therefore the mass enclosed within that sphere, from which we can attain the corresponding concentration $c_{200}$ via the relation shown above.
  
At this point it is important to highlight that $\beta$ is correlated with the parameter $\gamma$ (which quantifies the radial slope of the density profile of dark matter halos) and that the choice of density profile has some relation to the resulting $\beta$ \citep{Hansen2009}. However, the reason that we pick an Einasto profile over the NFW density profile (as well as whether or not it makes any difference to our results) will be made clear in the following section -- where we describe the particular method we utilize to derive velocity anisotropy profiles.

\section{Deriving velocity anisotropy profiles}
\label{sec:theory_exp}

Our strategy to derive the velocity anisotropy profile $\beta$ follows the long tradition of studies that have modeled the galaxy cluster as a collisionlesss system described by the anisotropic Jeans equation, reproduced here in Eq. \ref{eq:jeans} \citep{BinneyTremaine1987,BinneyMamon1982,Solanes1990}

If the mass of the system can be independently inferred, the potential-density pair can be defined and in combination with a second equation that relates the velocity dispersion of the system to its potential-density pair, the anisotropy profile $\beta$ can be derived by solving Eq. \ref{eq:jeans}. For instance, \cite{Natarajan97} --following \cite{BinneyMamon1982}-- reconstructs the radial velocity dispersion and use a weak lensing inference of the potential-density pair to derive $\beta$ for a single cluster. 

In contrast to this approach, we use a simple relation that relates the radial velocity dispersion to the line-of-sight velocity dispersion.  In particular, if cluster rotation is negligible, we have, $\langle v_{\theta}^2 \rangle = \langle v_{\phi}^2 \rangle = \langle v_{los}^2 \rangle$  \citep{diaferio1999}. Note that this holds for the velocities of the galaxies in the volume $dr^3$  centered on position $r$. And so from $\beta$ of  Eq.  \ref{eq:betasimple}, with  $\sigma_{t}^2 = \frac{1}{2} \big(\sigma_{\theta}^2 +\sigma_{\phi}^2 \big) $, we have that,

\begin{equation} 
\langle v_r ^2 \rangle = \frac{\langle v_{los}^2 \rangle}{1-\beta}.
\label{eq:diaferio}
\end{equation} 

This relation is utilized in the quite successful caustic mass technique (see \cite{diaferio1999, serra, geller2013}). For now, we take it as a given and we explore its validity at the end of this section.

In galaxy clusters, $\langle v_r ^2 \rangle = \sigma_r ^2 $, and so we have that with Eq. \ref{eq:diaferio}, the Jeans equation (Eq. \ref{eq:jeans}) now becomes,

\begin{equation} 
\frac{d}{dr} \bigg(\rho \frac{ \langle v_{los}^2 \rangle}{1-\beta}\bigg) + \frac{2\beta  \rho}{r}  \bigg( \frac{ \langle v_{los}^2 \rangle}{1-\beta}\bigg) =  -\rho \frac{d\phi}{dr}.
\label{eq:jeans2}
\end{equation} 
At first sight, solving this highly nonlinear differential equation for $\beta$ seems like a daunting task. However, after a bit of algebra we can re-write Eq. \ref{eq:jeans2} in the following way,
\begin{equation} 
\frac{d \beta}{dr} + (1-\beta) X + (1-\beta)^{2} Y + \beta (1-\beta) Z = 0 ,
\label{eq:jeans3}
\end{equation} 
where we have used the following redefinitions,

\begin{align} 
\begin{split} 
X \equiv &{\frac{1}{(\rho \sigma_{los}^2)}}\frac{d(\rho \sigma_{los}^2)}{dr},\\
Y\equiv & \frac{1}{ \sigma_{los}^2}  \frac{d\phi}{dr},\\
Z\equiv & \frac{2}{r}.
\end{split} 
\end{align} 

Now, redefining some variables again, $u \equiv 1-\beta$, we can re-write the  Eq. \ref{eq:jeans3} in the following form,
\begin{align} 
\begin{split} 
 \frac{du}{dr}   + P(r) u -  Q(r) u^2 = 0 .
\label{eq:bernoulli}
\end{split} 
\end{align} 
Where we have defined,
\begin{align} 
\begin{split} 
P(r) \equiv & - X - Z,\\
Q(r)\equiv & Y-Z. \\
\end{split} 
\end{align} 
The re-definitions and variable changes have allowed us to recast the Jeans equation (Eq. \ref{eq:jeans2}) as a Bernoulli differential equation (Eq. \ref{eq:bernoulli}) which has a well-known analytic solution. For our specific functions, after some algebra and after solving for the integrating factor, solving Eq. \ref{eq:bernoulli} yields,
\begin{equation} 
\beta = 1 -  \frac{  \rho \sigma_{los}^2 r^2 }{  I_1 - I_2  }.
\label{eq:betar}
\end{equation} 
Where the integrals are given by, 

\begin{align} 
\begin{split} 
 I_1(r) = \int_{r_1}^{r}  2\rho \sigma_{los}^2 r'   dr'   ,\\
I_2(r) = \int_{r_1}^{r}   \rho \frac{d\phi}{dr'} r'^2  dr',
\label{eq:int}
 \end{split} 
\end{align} 
and where the lower integration limit $r_1 = 0.05$ Mpc. We explain the reason choosing this integration limit below.

As such, the radial anisotropy profile for a given cluster can be attained with Eqs. \ref{eq:betar}-\ref{eq:int} for an independently inferred mass profile which determines the potential-density pair (that is, $\phi$ and $\rho$, respectively) and a measurement of the line-of-sight velocity profile of galaxies ($\sigma_{los}$).

As anticipated in the previous section, for the potential-density pair we pick the Einasto model.  From the Einasto density profile (Eq. \ref{eq:einasto_den}) we can derive the gravitational potential $\phi$ using the integral form of the Poisson equation \citep{retana},

\begin{equation}        
\phi(r) = -\frac{{\rm GM}}{r} \Bigg{[} 1 - \frac{\Gamma\big{(}3n,\big(\frac{r}{r_0}\big)^{1/n}\big{)}}{\Gamma(3n)} + \frac{r}{r_0}\frac{\Gamma\big{(}2n,\big(\frac{r}{r_0}\big)^{1/n}\big{)}}{\Gamma(3n)}\Bigg{].} \label{eq:einasto_pot}
\end{equation}
Note that we calculate the derivative of $\phi$ in  Eqs. \ref{eq:betar}-\ref{eq:int} numerically. 

At this point it is important to clarify the reason we chose to use an Einasto rather than an NFW potential-density pair in our analysis and how this relates to why we pick $r_1 = 0.05$ Mpc. 

As shown in both cosmological N-body simulations \citep{miller2016} and observational data \citep{Stark2016ApJ} the gravitational potential inferred from the NFW density profile yields a profile that is biased high from what is expected. In particular, the NFW density is significantly steeper than the Einasto density as  $ r \rightarrow 0$, which also means that the potential ($\phi$) and its derivative ($d\phi/dr$) are higher for the NFW model. As such, when Eqs. \ref{eq:betar}-\ref{eq:int} are integrated, we expect to see a difference between the NFW and Einasto profiles in the resulting $\beta.$ Because we want our results to be as model-independent as possible, and because there is still considerable uncertainty in the modeling of galaxy cluster cores, the lower limit ($r_1$) of the integrals in Eq. \ref{eq:int} is chosen to be 0.05 Mpc. Note that we cut off the derived $\beta$'s at $R_{200}$, where the Einasto and NFW models agree with each other -- though this upper limit to the extrapolated radial range does not make a difference to our results. As such when the radial range used in the integrands of Eq. \ref{eq:int} is $ 0.05 \leq R \leq  R_{200}$ Mpc,  we  see no significant difference between the Einasto and NFW models. This is due to the fact that the major difference between these models arises from the modeling of the core of galaxy clusters and its outskirts (that is, $R > R_{200}$). While we find the model dependence of the inferred $\beta$ to be a crucial matter that merits further study, we relegate a more thorough analysis of it to a future investigation.

Secondly, we also want to note that in contrast to the results of \cite{miller2016} and \cite{Stark2016ApJ}, our Jeans equation is not dependent on cosmology. The reason for this is that while those two studies work with observables that are directly proportional to the potential (and must therefore take into account the specific normalization of the potential) the Jeans equation is only dependent on the spatial derivative of the potential for which the effects of the expansion of the Universe can be neglected. This is known as the ``Jeans swindle'' and has been formally justified in \cite{JeansSwindle}.

Thirdly, another important aspect to highlight about our approach is that unlike the common and widely-used method developed in \cite{Solanes1990}, our approach allows us to straightforwardly calculate the uncertainties of $\beta$.  As noted by both  \cite{macs1206} and \cite{BivianoKatgert2004}, the derived uncertainties on $\beta$  are very difficult to calculate with the approach of \cite{Solanes1990} because it works through a complicated set of coupled differential equations. This is not the case for our method given that we solve for $\beta$ analytically by recasting it as a Bernoulli differential equation. In particular, for our simulated clusters, we take into account the uncertainty in measuring the line-of-sight velocities simply by creating an array of $\beta$'s which correspond the $1\sigma$ errors on the measured line-of-sight velocity profiles ($\sigma_{los}$). For our real clusters, we consider both the error on the inferred $\sigma_{los}$ profiles as well as the error on the mass profiles (which propagates to $d \phi/dr$ and $\rho$) simply by recalculating $\beta$ for the range of uncertainty in either the mass and the line-of-sight dispersion profiles.

Lastly, as mentioned before, our approach is based on the relation shown in Eq. \ref{eq:diaferio} which is utilized in the well-established caustic  mass technique \citep{diaferiogeller1997,diaferio1999, serra, geller2013}. Nonetheless, we want to confirm the validity of Eq. \ref{eq:diaferio}. In particular, we want to analyse how $\beta^{los}$ compares to the directly measured $\beta$ from galaxies in simulated halos via Eq. \ref{eq:betasimple}, that is, the  ``true'' $\beta$. To that end, we define Eq. \ref{eq:diaferio} to be the "line-of-sight" velocity anisotropy profile ($\beta^{los}$),

\begin{equation} 
\beta^{los} = 1 - \frac{\sigma_{los}^2}{\sigma_{r}^2}.
\label{eq:betavlos}
\end{equation} 

\begin{figure}
\epsscale{1.21}
\plotone{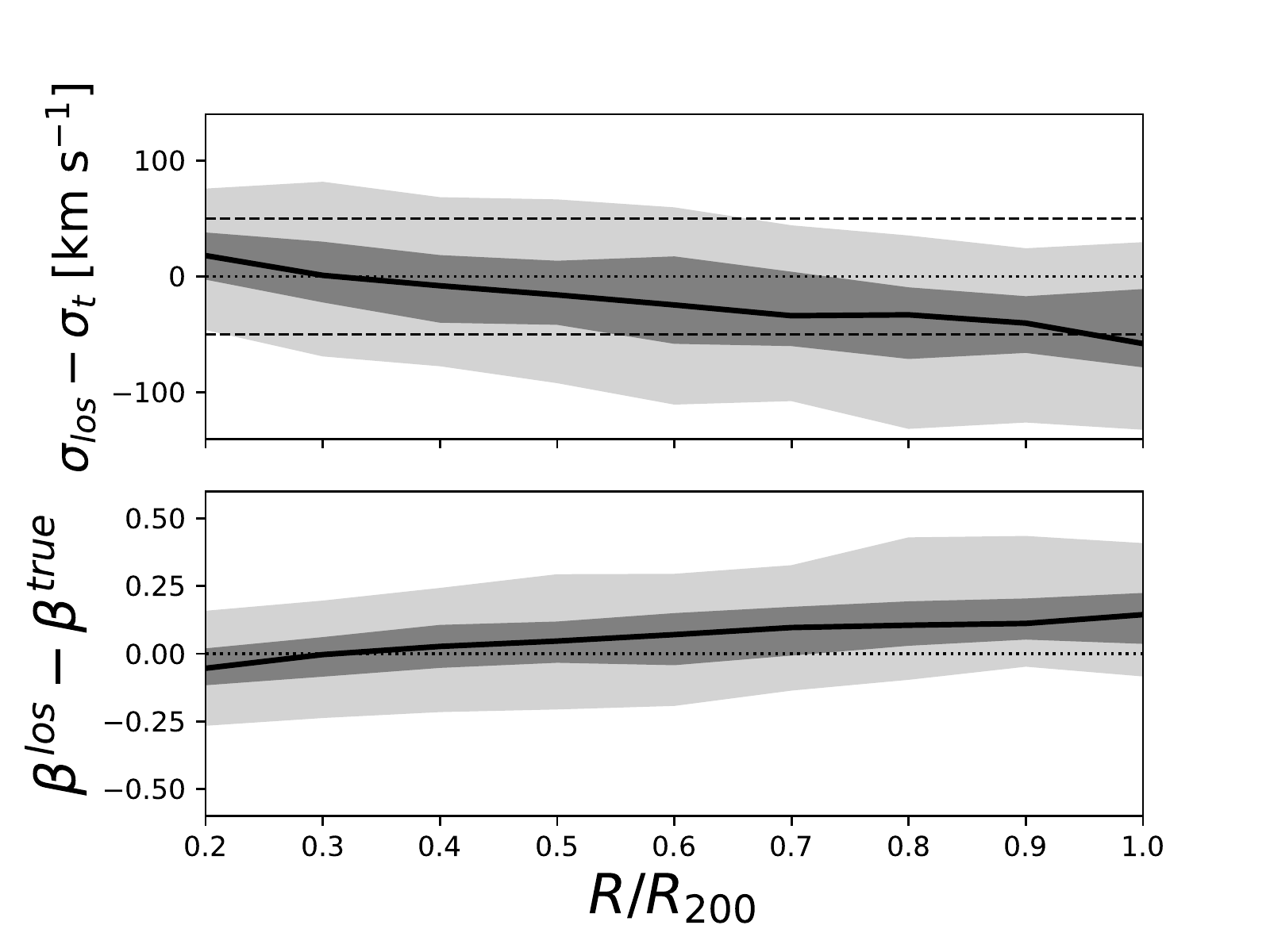}
\caption{ \textit{Top panel:} Difference between the measured line-of-sight velocity dispersion profile ($\sigma_{los}$) and the tangential velocity dispersion ($\sigma_t$) for the 100 synthetic clusters from  cosmological N-body simulations. The horizontal dashed lines represent the 1$\sigma$ error on our measurement of $\sigma_{los}$, $\sim 50$ km s$^{-1}$.\textit{ Bottom panel:} Difference between the ``true'' $\beta$ measured via Eq. \ref{eq:betasimple} and the ``line-of-sight'' $\beta$ (Eq. \ref{eq:betavlos}). See text for details. The black line shows the median (50th percentile) and the dark (light) gray bands represents the 68th (90th) percentile. The small bias in $\sigma_{los}$ leads to a systematic bias in the inferred $\beta$.} \label{fig1}\end{figure}

The difference between $\beta^{los}$ and the ``true'' $\beta$ is shown in Fig. \ref{fig1}. In the top-panel, we show that while on average the assumption made by in the caustic technique yields a small bias, the relation between the line-of-sight velocity dispersion ($\sigma_{los}$) and the transversal velocity dispersion ($\sigma_{t}$) depends on radius. We note that, while this systematic is still within our overall uncertainties, it must be corrected for. As shown in the bottom-panel of the same figure, if this systematic is not corrected for, our technique can yield an average $\beta$ that is higher from the true value by a difference of 0.14 at $R_{200}$. We also note that while this systematic is important, it still lies within our current uncertainties in the measurement of $\sigma_{los}$, which are about $50$ km s$^{-1}$ (see the horizontal dashed lines on the top panel of Fig. \ref{fig1}). We note that no significant mass-concentration dependence is found on the results of Fig. \ref{fig1}.

Having checked the validity of our presuppositions and having specified our approach to derive $\beta$ profiles, we now describe our results.

\section{Results}
\label{sec:results}

Using Eq. \ref{eq:betar} (see Section \ref{sec:theory_exp}) we test our approach with the 100 synthetic clusters from cosmological N-body simulations and then move on to derive $\beta$ profiles for archival data of 35 galaxy clusters. 

\subsection{Synthetic data results}
\label{sec:syntheticdataresults}
\begin{figure}
\epsscale{1.21}
\plotone{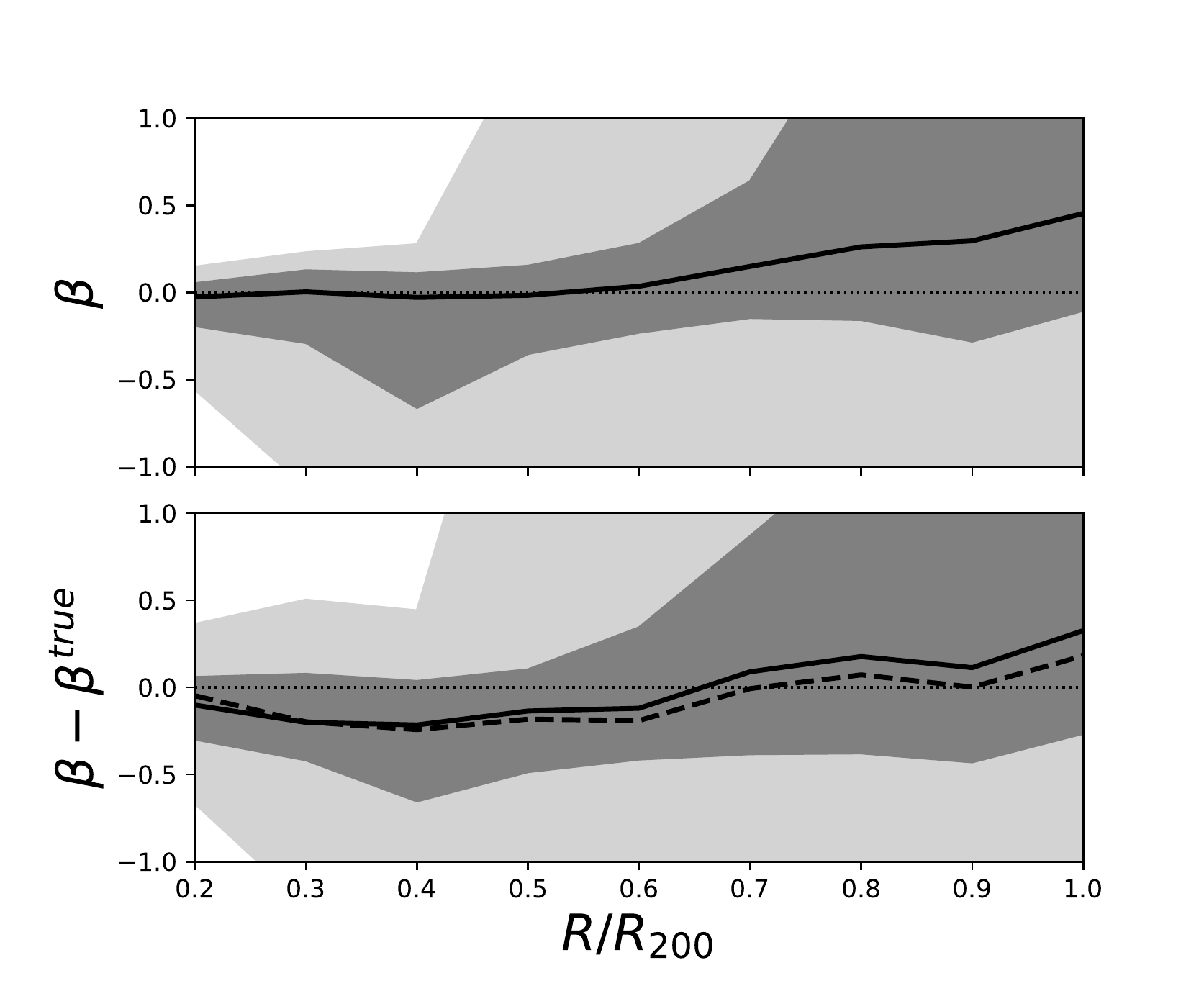}
\caption{Results for the sample of 100 synthetic clusters from cosmological N-body simulations. \textit{Top panel:} median (black line) with 68\% and 90\% percentile bands (dark and light gray, respectively).  $\beta$ is calculated with Equation \ref{eq:betar} (see Section \ref{sec:theory_exp}) and the resulting median is $\beta \sim 0.1$. \textit{Bottom panel:} The difference, $(\beta - \beta^{true})$, between $\beta$ as calculated with our approach and the ``true" anisotropy parameter ($\beta^{true}$) calculated directly from the simulated galaxies via Eq. \ref{eq:betasimple}. The dark and light gray bands represent the same percentiles as those shown in the top panel. Note that the fractional difference between the ``true" $\beta$ and our inference is about zero. The dashed black line is the resulting difference after having corrected for the systematic shown in the bottom panel of Figure \ref{fig1}. We note that while Eq. \ref{eq:betar} can yield profiles in which $\beta > 1$, we require that the solutions yield physical results. Section \ref{sec:syntheticdataresults} for more details.}
 \label{fig2}\end{figure}

The top panel of Figure \ref{fig2} shows our results for the 100 synthetic clusters using Equation \ref{eq:betar}. The bands represent the 68th (dark gray) and 90th (light gray) percentiles of the sample.  The median (black line) hovers around $\beta \sim 0.1 $. In the lower panel of Figure \ref{fig2}, we plot up the difference between $\beta$ as estimated through our approach (Eq. \ref{eq:betar}) and the ``true'' $\beta$ estimated directly from Eq. \ref{eq:betasimple} (solid line). Figure \ref{fig2} demonstrates that our approach works with accuracy and precision once radially averaged, $\langle \beta - \beta^{true} \rangle \sim 0 $. The dashed black line in the bottom-panel shows the resulting difference after having corrected for the systematic shown in the bottom panel of Figure \ref{fig1}. This test also allows us to set a baseline for the systematic uncertainty, which incorporates both the observational systematic (see Figure \ref{fig1}) as well as any additional systematic errors introduced by the technique itself. Using Figure \ref{fig1}, we define the overall systematic error on the mean measurement of $\beta$ to be $\sim \pm 0.15$, which is the maximum deviation between our inference of $\beta$ and the true $\beta$.

We note that the result of Fig. \ref{fig2} does not take into account the uncertainty on $\sigma_{los}$ nor the weak lensing inferred density and potential that exist in the real data. When doing so, one can find $\beta$ profiles that occasionally exceed the physical condition from Equation \ref{eq:betasimple} that $\beta < 1$. This is due to the fact that while the integrals of Equation \ref{eq:int} are well-behaved in the radial range discussed in the previous section, the solutions to Eq. \ref{eq:betar} can still yield non-physical results for any given combination of mass ($\rho$ and $d\phi/dr$) and dynamics ($\sigma_{los}$). That is, for some clusters, there is a combination of dynamics and weak lensing that yields a $\beta$ profile that is unphysical within the uncertainties on $\sigma_{los}$ and the weak lensing mass profile estimates. In particular, the only case which can yield a non-physical result (namely, $\beta > 1$) occurs whenever $I_1 > I_2$ at some given radius. As such, when we include errors on the line-of-sight dispersions and the weak lensing masses, and in contrast to what is shown on Figure \ref{fig2}, we require that the solutions yield physical results ($\beta < 1$). Specifically, we do not use $\beta$ profiles that yield $\beta > 1$ within the virial sphere (that is, if $\beta(R\leq R_{200}) > 1$).

\subsection{Archival data cluster results}

Having shown our approach to derive $\beta$ works well in N-body simulations, and in particular, that it allows us to recover the true $\beta$ to within $\pm 0.15$, we now derive the $\beta$ profiles for the 35 clusters from the archival data. 

\begin{figure}
\epsscale{1.21}
\plotone{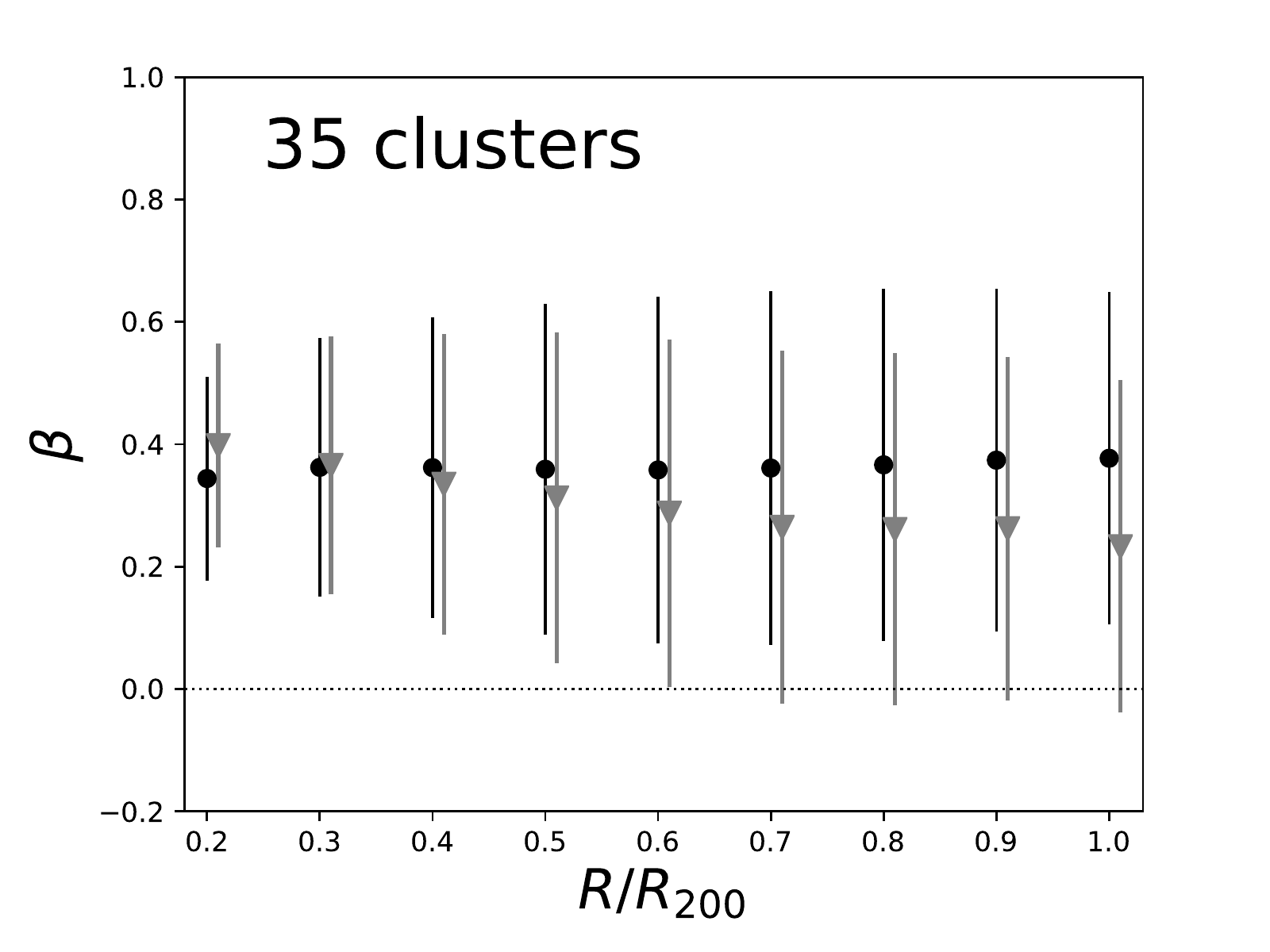}
\caption{Resulting $\beta$ profile for the 35 clusters of the  archival data. The black dots are the weighted mean of each individual profile calculated with Eq. \ref{eq:betar} 
(see Section \ref{sec:theory_exp}) and the error is the 1$\sigma$ uncertainty on the weighted mean. Note that each individual profile that is averaged here contains both the uncertainty in mass and line-of-sight velocity dispersion. The global value of $\beta$ averaged between 0.2 and $R_{200}$ is $\langle \beta( 0.2 \leq R/R_{200} \leq 1 )\rangle = 0.35 \pm 0.28 $. This means that the galactic orbits are mostly radially anisotropic. The gray triangles are the result of correcting for the systematic shown in the bottom-panel of Fig. \ref{fig1}. We note that the overall effect still produces a result that lies within our 1$\sigma$ error. Note that we have slightly shifted the gray triangles to higher radii simply to make the plot legible.}
 \label{fig3}\end{figure}

In Fig. \ref{fig3} we plot the weighted average of the 35 $\beta$ profiles. More specifically, we take the average $\beta$ at each radial bin, now divided in increments of $\Delta (R/R_{200}) = 0.1 $, by weighing the individual cluster profiles-all of which have different uncertainties at a given radii incurred from both their respective line-of-sight velocity dispersion profile and their mass profile uncertainties. The result is shown in the black dots with the $1\sigma$ error on the mean in Fig. \ref{fig3}. The gray triangles are the resulting weighted average (and respective $1\sigma$ error on the mean) when we correct for the systematic shown in Fig. \ref{fig2} by subtracting off the difference (see bottom-panel of that same figure) from our weighted average. We note that the resulting weighted average after this systematic is included as well within the $1\sigma$ error.

The radially averaged $\beta$, shown with black dots in Fig. \ref{fig3}, yields a global value of $\langle \beta \rangle = 0.35 \pm 0.28$. The grey triangles in Fig. \ref{fig3}, yield a global value of $\langle \beta \rangle = 0.26 \pm 0.28$. These  results imply  that the observed velocity anisotropy profile of galaxy clusters is radially anisotropic. We discuss the implications of this result and compare it to other studies in the next section.

\begin{figure*}
\epsscale{1.2}
\plotone{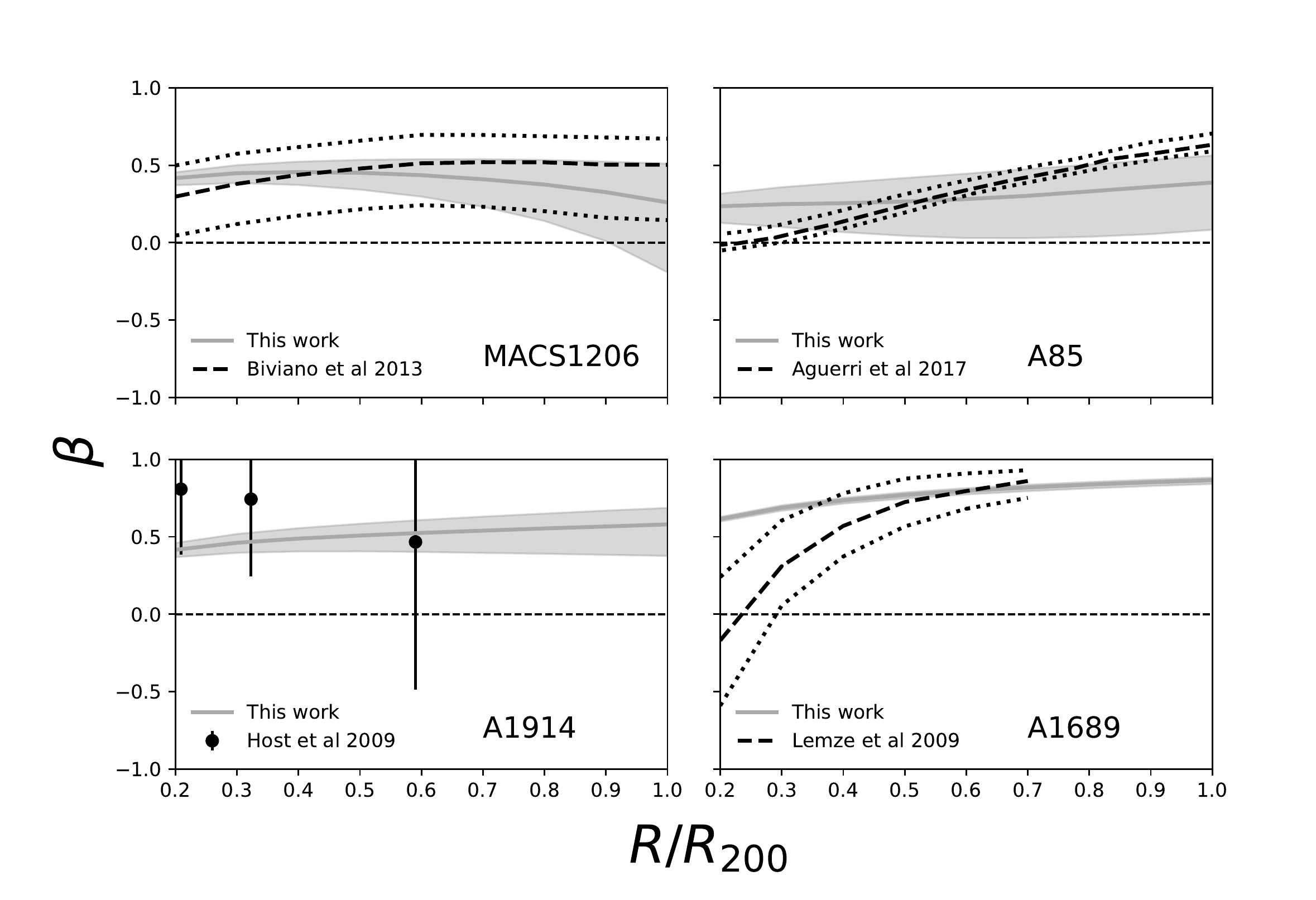}
\caption{ Comparison of 4 of the 35 individual $\beta$ profiles with results from the literature. For all the panels, the gray band represents the 68th percentile error on $\beta$ from both the uncertainty in the weak lensing mass profiles as well as the uncertainty in the line-of-sight velocity dispersion profiles. See text for details.}
 \label{fig4}\end{figure*}

\section{discussion and summary}
\label{sec:discussion}

The goal of this work is to constrain the average radial $\beta$ profile of clusters, $\langle \beta \rangle$, using a set of clusters with combined weak lensing and dynamical phase space data. In what follows we compare our results to other numerical studies as well as other observational data results. 

Individually, many of the clusters have large errors in their inferred $\beta$ profiles. However, there are a few clusters with well-constrained mass profiles and well-sampled phase spaces which also have previously published anisotropy profiles. As such, we compare our results for these clusters. In Figure \ref{fig4}, we plot four of these cluster $\beta$ profiles as well as the profiles derived by \cite{macs1206} (upper-left panel), \cite{Aguerri2017} (upper-right panel),
\cite{lemze2009} (lower-right panel) and \cite{host2008} (lower-left panel). All of these papers model the relation between dynamics, anisotropy, and the mass  of galaxy clusters via the Jeans equation, although they utilize different methods to estimate the mass profile. For instance, \cite{macs1206,Aguerri2017,lemze2009} all combine different mass profile estimates, while \cite{host2008} exclusively uses X-ray mass profiles. For all cases, in Fig. \ref{fig4}, the $\beta$ we derived is shown in a gray band that represents the 68th percentile uncertainty that takes into account the uncertainty in both the line-of-sight velocity dispersion profile and the weak lensing mass profile of each respective cluster. In black (in dots or lines) we show either the median or average along with the $1\sigma$ error on $\beta$ from each paper just cited. First of all, note that in Fig. \ref{fig4} we do not apply the aforementioned correction factor given that it is only valid on average. Nonetheless, note that in all cases our results agree, within respective uncertainties, with the previously published results. That is, galaxy orbits tend to be more  radially anisotropic the farther away from the core of the cluster we go. However, our results seem to disagree with both \cite{lemze2009}'s and \cite{Aguerri2017}'s results at smaller radii. In particular, our profiles of  A1698 and A85 do not become isotropic as quickly as expected by these two other results.

For the case of A1689, we note that \cite{lemze2009} uses a very high concentration ($c_{200} > 10$), which is $\sim$3.3 times larger than the concentration we use. The relation between concentration and $\beta$ was studied in \cite{wojtak}. Their conclusion is the same as ours: a lower concentration yields a higher $\beta$.

Now, for the case of A85, while we use a very similar mass and concentration as what is used in \cite{Aguerri2017}, we find that the central velocity dispersion parameter we fit is smaller than what is found by \cite{Aguerri2017} using all galaxies (that is, as opposed to using cuts in color or luminosity which can also bias the calculated line-of-sight of sight velocity dispersion). Note that a lower line-of-sight dispersion also yields a higher $\beta.$ We want to highlight, however, that our $\beta$ profile of A85 agrees with what is derived by \cite{Hwang} (see Fig. 19, top-left panel for cluster A85, within $R_{200}$) but we do not overplot their results because their scatter is too large.

\begin{figure}
\epsscale{1.2}
\plotone{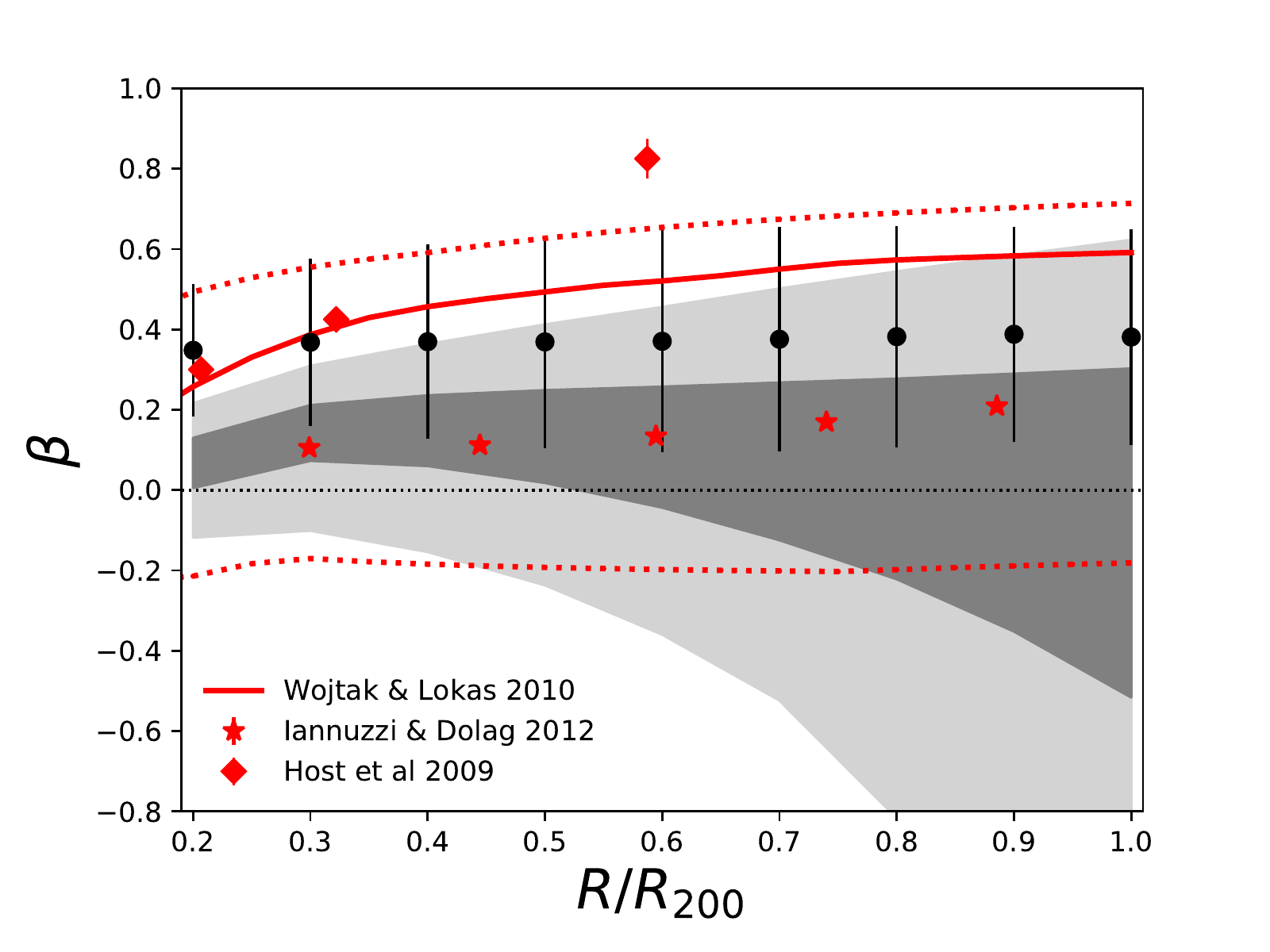}
\caption{Comparison between $\beta$ profiles from the literature and our work. The $\beta$ calculated with Eq. \ref{eq:betar} (see Section \ref{sec:theory_exp})  from 100 N-body simulated clusters is shown in light/dark gray bands  and the $\beta$ from the 35 archival data clusters is shown with black dots (same as in Fig. \ref{fig3}). In contrast to those of Fig. \ref{fig2}, the synthetic cluster results (grey bands) now take into account the uncertainty in the line-of-sight velocities as well as consider only physical results ($\beta \leq 1$). In red diamonds, we show the results of the median and $1\sigma$ error on $\beta$ calculated for 16 clusters by \cite{host2008}. In the red and dotted lines, we show the $\beta$ profiles calculated by \cite{wojtak} for their sample. Lastly, the red stars are the averaged $\beta$ for 1000 simulated clusters by \cite{iannuzzi}. Note that the $1\sigma$ error for the  \cite{iannuzzi} result is too small to show. The black dots are the same as those of Fig. \ref{fig3}.}
 \label{fig5}\end{figure}

In Figure \ref{fig5}, we compare our radially averaged anisotropy profile to other average profiles from the literature. In this case, we only show the version that has not been corrected for the systematic bias identified using the simulations (i.e., we show the black dots from Figure \ref{fig3}). We only compare our results with previously published average profiles of $\beta$ based on samples of clusters (i.e., not individual systems). We also compare our resulting $\beta$ with the Millennium simulation results of \cite{iannuzzi}. Note that we do not compare our results with the work of authors who assume a constant $\beta$ or specify a $\beta$ only for a specific class of galaxies (such as that of \cite{BivianoKatgert2004}).

Our average radial anisotropy profile  agrees well with the average profile of 16 clusters derived by \cite{host2008} (see Fig. 7 of that paper). In particular, \cite{host2008} also find a median value of $\beta$ that implies radial anisotropy, such that $\beta(R/R_{200} \geq 0.2) >  0.3$. We  overplot the median $\beta$ calculated by \cite{host2008} in red  diamonds, alongside our results, on Fig. \ref{fig5}.

We compare our results with the $\beta$ profile of \cite{wojtak} in Fig. \ref{fig5}. The red (solid and dotted lines) represent the quartiles of the  best-fit $\beta$ as a function of $R/R_{200}$.

Lastky, we also compare our results with our aforementioned 100 synthetic clusters from N-body simulations, also shown in Fig. \ref{fig5}. In particular, in contrast to the results of Fig. \ref{fig2}, we now consider the uncertainties in the measured $\sigma_{los}$ (and its fits) from the synthetic cluster sample which is shown in the light gray (dark grey) bands of Fig. \ref{fig5} which represent the 68th (90th) percentiles. Finally, we make a further comparison with simulations by comparing to the independent analysis of the $\beta$ profile (from the 3D data) measured by \cite{iannuzzi}, who uses 1000 synthetic galaxy clusters. They find  $\langle \beta \rangle = 0.253 \pm 0.01$ (their profile is shown with red stars in Figure \ref{fig5}). Note that we find that our simulation results (shown in the gray bands) agree well with the simulation results of  \cite{iannuzzi}.

Considering that we are using the same method for both the data and the simulations, we find good agreement after we allow for systematic errors in the data. We note that the synthetic cluster sample is much lower in redshift than the real data sample. More specifically, the synthetic sample has a redshift range that is $z_c \leq 0.15$ , whereas the cluster redshifts for the real data is $\langle z_c \rangle = 0.25$.  As investigated in N-body simulations by \cite{iannuzzi}, we do not expect an appreciable redshift evolution of the global $\beta$ parameter.
We also note that our sample of 35 real data clusters have an average mass that is 4.4 times higher than the average mass of the 100 synthetic clusters. Recalling the negative correlation between mass and concentration, we might expect that a lower concentration (higher mass) yield a higher $\beta$ for the data compared to the simulations. Also, recall that our fits to the line-of-sight velocity dispersion profiles are biased low with respect to the measured profile by $\sim 43$ km s$^{-1}$, which would bias our $\beta$'s high.

Finally, the extrapolation of functions such as $\sigma_{los}$ and the density profiles well into the core and the outskirts of galaxy clusters is definitely a weakness in this technique and others like it which depend on integration over the entire radial profiles of clusters (i.e., most Jeans-like analyses). We attempted to mitigate the effects of uncertainties in the cores of the clusters by integrating from 0.05 Mpc outward. However, this deserves further study. Furthermore, better fitting functions for $\sigma_{los}$ as well as better-sampled clusters also yield better fits to the line-of-sight velocity dispersion. This could partially explain why the 35 real clusters sit on the upper end of the simulation results (compare black dots and gray bands in Fig. \ref{fig5}). Nonetheless, we emphasize that this slight disagreement is within the total systematic and statistical error we estimate. 

Our work highlights a new advance in the measure of the velocity anisotropy of galaxy clusters. First, we develop a novel technique to combine the weak lensing mass information about the cluster mass profile with the observed line-of-sight velocity dispersion profile. Second, we test the accuracy and precision of this technique to realistic simulations, where the phase space data are projected onto the line-of-sight. Finally, we apply the technique to a sample of tens of clusters that have the required data. We find that within observed errors, the simulations and the data agree quite well. In particular, we find that the cluster velocity anistropy profile is flat with a value $\langle \beta \rangle = 0.35 \pm{0.28}$ (stat) $\pm{0.15}$ (sys) implying that galactic orbits tend to be radially anisotropic.

Lastly, we note that our Jeans analysis approach requires that galaxies are distributed like the underlying diffuse matter. As such, instead of using the weak lensing matter density profiles we could have instead opted to use the number density of the tracer galaxies.  However, there is significantly more scatter in the galaxy density profiles due to the poor overall sampling of the phase-spaces (from about fifty to a few hundred as noted in Section 2.2). Furthermore, in contrast to  the weak lensing density profile error estimation technique, we are unable to rigorously assess the errors in the galaxy density profiles.  The reason why we cannot rigorously measure the error bars of the galaxy density profile is that this would entail incorporating 3-dimensional membership, a local background model, as well as targeting selection for the archival data \citep{Munari2014}.  This analysis is therefore beyond the scope of this paper. We defer a more thorough analysis of galaxy tracer populations to a future study. Nonetheless, this is a crucial aspect of our approach. Unlike other approaches to derive $\beta$'s \citep{Solanes1990}, our analytic solution allows us to straightforwardly calculate the uncertainties on the anisotropy parameter -- folding through the both the uncertainties in the the velocity dispersion profiles and weak lensing masses.


\section{acknowledgments}
AS  is  supported  by  the  National  Science Foundation under Grant No.  1311820. CJM and VH are supported by the Department  of  Energy  grant  de-sc0013520. We thank Radoslaw Wojtak for sending us the $\beta$ profile used in Figure 5. The Millennium Simulation databases used in this paper and the web application providing online access to them were constructed as part of the activities of the German Astrophysical Virtual Observatory (GAVO). This research has made use of the VizieR catalogue access tool, CDS, Strasbourg, France.
\bibliographystyle{apj}
\bibliography{stark}
\newpage

\end{document}